\documentclass[11pt]{article}
\usepackage{amsfonts,amssymb}

\def\squareforqed{\hbox{\rlap{$\sqcap$}$\sqcup$}}
\def\qed{\ifmmode\squareforqed\else{\unskip\nobreak\hfil
\penalty50\hskip1em\null\nobreak\hfil\squareforqed
\parfillskip=0pt\finalhyphendemerits=0\endgraf}\fi}
\newtheorem{theorem}{Theorem}
\newenvironment{proof}{\begin{trivlist}\item[]{\flushleft\bf Proof }}
{\qed\end{trivlist}}

\DeclareMathSymbol{\itTheta}{\mathalpha}{letters}{"02}
\DeclareMathSymbol{\leqslant}{\mathrel}{AMSa}{"36}  

\newcommand{\subgroup}{\leqslant}        % subgroup symbol
\newcommand{\Co}{{\mathbb C}}            % the field of complex numbers
\newcommand{\ket}[1]{\mbox{$| #1 \rangle$}}

\title{A Quantum Observable for the\\ Graph Isomorphism Problem}

\author{Mark Ettinger\,%
\thanks{\,\mbox{NIS--8}, \mbox{MS~B230}, 
Los Alamos National Laboratory, Los Alamos, NM~87545, USA.
Email: \texttt{\boldmath ettinger$\mathchar"40$lanl.gov}.}\\%
{\protect\small\sl Los Alamos National Laboratory\/}%
\and
Peter H{\o}yer\,%
\thanks{\,BRICS, Department of Computer Science, 
University of Aarhus, \mbox{DK--8000} \mbox{Aarhus~C}, Denmark.
Email: \texttt{\boldmath hoyer$\mathchar"40$brics.dk}.}\\%
{\protect\small\sl BRICS\/}\,%
\thanks{\,Basic Research in Computer Science, 
Centre of the Danish National Research Foundation.}}

\date{} %{January 1999}

\begin{document}

\maketitle

\begin{abstract}
Suppose we are given two graphs on $n$ vertices.
We define an observable in the Hilbert 
space $\Co[(S_n \wr S_2)^m]$
which returns the answer ``yes'' with certainty if the graphs are 
isomorphic and ``no'' with probability at least $1-\frac{n!}{2^m}$ 
if the graphs are not isomorphic.  
We do not know if this observable is efficiently 
implementable. 
\end{abstract}

\section{Introduction}
The graph isomorphism problem is to determine if two graphs 
$\Gamma_1, \Gamma_2$ on $n$ vertices are isomorphic.   
Let $\Gamma$ be the disjoint union graph of $\Gamma_1$ and $\Gamma_2$.  
Without loss of generality we may assume that both 
$\Gamma_1$ and $\Gamma_2$ are connected.  In~this case the automorphism
group of $\Gamma$ is a subgroup of the wreath product 
$S_n \wr S_2$ (which is itself a subgroup of $S_{2n}$).
Clearly, knowledge of a set of generators for this automorphism group
is sufficient to decide the isomorphism question.  This fact has
resulted in the suggestion that a quantum computer may be able to
efficiently find a set of generators for the automophism group and
thus solve the graph isomorphism problem.  This idea originates in the
hidden subgroup view of quantum algorithms~\cite{EH99}.  
The Abelian hidden subgroup problem can be solved 
in polynomial time and utilizes the 
Fourier observable or, equivalently stated, the quantum algorithm
utilizes the quantum Fourier transform.  We use the terminology ``Fourier 
observable'' to emphasize the particular point of view germane to the main
result of this paper.  A~quantum algorithm is simply 
a unitary change-of-basis
transformation from the computational basis to the basis of the observable.
We remark that in this paper ``Fourier observable'' refers 
to the Abelian case.  The difficulties of
finding hidden subgroups of noncommutative groups have been explored in
several papers including~\cite{EH99,RB98}.  
For more information on the Abelian hidden subgroup problem,
see for example the references in~\cite{EH99,RB98}.

There are several important differences between the observable
presented here and the Fourier observable.  The first difference is
that the present observable operates on a larger Hilbert space.
Recently it was shown in~\cite{EHK99} that a hidden 
noncommutative group may be found in only polynomially many calls to
the oracle function, although the algorithm given in~\cite{EHK99} 
requires exponential time.  This result was proved by showing that the 
{\em tensor product states\/} 
corresponding to different possible hidden subgroups are almost
orthogonal in the larger Hilbert space $\Co[G^m]$.  In the present paper
we work in such a Hilbert space.  The second difference is that our
observable reveals nothing {\em directly\/} about the automorphism group
other than whether or not it contains an isomorphism between the two
graphs.  However we may then find the full automorphism group using a
well known classical reduction~\cite{Mathon79}.
Thirdly and finally, whereas it is known that the Fourier observable is
efficiently implementable, we have not been able to demonstrate this
for the observable presented below.  Such an efficient implementation
would result in a polynomial-time quantum algorithm for the graph
isomorphism problem.

\section{The Observable}
Let $G = S_n \wr S_2$.  
Since the wreath product is a semidirect
product $(S_n \times S_n) \rtimes S_2$ we write an element as a
triple $(\sigma,\tau,b)$.  We refer to any element of $G$ of the form
$k = (g,g^{-1},1)$ as an {\em involutive swap}.  
Let $\mathcal{H} = \Co[G^m]$.
Note that $\dim(\mathcal{H}) = |G|^m = 2^m (n!)^{2m}$.  
For each $k \in G$, we define a
{\em $k$-vector\/} to be a vector of the form:
$$ \frac{1}{\sqrt{2^m}}\Big((\ket{c_1} + \ket{c_1k}) \otimes
\cdots \otimes (\ket{c_m} + \ket{c_mk})\Big)$$
for some $c_1,\dots,c_m \in G$.  
Define $\mathcal{H}(k)$ to be the subspace
spanned by all $k$-vectors.  Notice that if $v_1$ and $v_2$ are unequal
$k$-vectors then they are orthogonal.  
Therefore $\dim(\mathcal{H}(k)) =  (\frac{|G|}{2})^m$.  
Let $\mathcal{H}_1 = \sum_k \mathcal{H}(k)$ be
the sum over all $n!$ involutive swaps.  
Notice that $\dim(\mathcal{H}_1) \leq n!(\frac{|G|}{2})^m$.
Let $\mathcal{H}_0$ be the
orthogonal complement to $\mathcal{H}_1$ in $\mathcal{H}$.
Our observable is defined as $L = \lambda_0P_0 + \lambda_1P_1$ where
$P_0$ and $P_1$ are projections onto $\mathcal{H}_0$ and
$\mathcal{H}_1$ respectively, and $\lambda_0, \lambda_1 \in \Co$.  

Let us see what this observable yields when we apply it to the states
that we may easily produce, i.e., tensor products of coset states.
Let $H \subgroup G$ be the automorphism
group of $\Gamma$.  Let $\ket{\psi}$ be a tensor product of coset
states of $H$, i.e.{}
$$\ket{\psi} = \ket{c_1H} \otimes \cdots \otimes \ket{c_mH},$$ 
where for any non-empty subset $X \subseteq G$, 
$$\ket{X} = \frac{1}{\sqrt{|X|}} \sum_{x \in X} \ket{x}.$$

\begin{theorem}
If $\Gamma_1$ and $\Gamma_2$ are isomorphic then 
$\langle \psi |P_1| \psi \rangle = 1$.
\end{theorem}

\begin{proof}
If $\Gamma_1$ and $\Gamma_2$ are isomorphic via the involutive swap~$k$ 
then $k \in H$, 
and thus any coset state of~$H$ may be written (omitting normalizations):
$$\ket{cH} = \ket{ch_1} + \ket{ch_1k} + \cdots
+\ket{ch_{\frac{|H|}{2}}} + \ket{ch_{\frac{|H|}{2}}k}.$$  It is then
easy to see that tensor products of these cosets state can be written
as sums of $k$-vectors.  For example $$\ket{c_1H} \otimes \ket{c_2H} = 
(\ket{c_1h_1} + \ket{c_1h_1k}) \otimes (\ket{c_2h_1} + \ket{c_2h_1k})
+ \cdots.$$  
Any sum of $k$-vectors is, by definition, in
$\mathcal{H}_1$ and the result follows.
\end{proof}

\begin{theorem}
If $\Gamma_1$ and $\Gamma_2$ are not isomorphic then 
$\langle \psi|P_0| \psi \rangle \geq 1 - \frac{n!}{2^m}$.
\end{theorem}

\begin{proof}
Assume the graphs are nonisomorphic.
We show $\langle \psi|P_1| \psi \rangle \leq \frac{n!}{2^m}$.  
First, suppose 
$\ket{\psi} = \ket{g_1} \otimes \cdots \otimes \ket{g_m}
 = \ket{(g_1,g_2,\dots,g_m)}$.  
This occurs when both graphs are rigid and $H$ is trivial.  
For each involutive swap $k$ there exists exactly one $k$-vector
which is not orthogonal to $\ket{\psi}$ and this $k$-vector has the
form:
$$\frac{1}{\sqrt{2^m}}\Big(\ket{(g_1,\dots,g_m)} 
 + \ket{(g_1k,\dots,g_m)} + \cdots + \ket{(g_1k,\dots,g_mk)}\Big).$$
Therefore $\langle\psi |P(k)|\psi\rangle = \frac{1}{2^m}$, 
where $P(k)$ is the projection onto $\mathcal{H}(k)$.  This implies
$\langle\psi|P_1|\psi\rangle \leq \frac{n!}{2^m}$.  For
nontrivial $H$ the argument is almost identical except that since
$\ket{\psi}$ is not a basis state we must sum the probability
contributions over the support, resulting in identical conclusions.
\end{proof}

\section{Conclusion}
We have described a quantum observable on a Hilbert space 
for which the logarithm of its dimension is
polynomial in the number of vertices of the graphs.  
This observable decides the isomorphism question with high probability.  
However we do
not know if this observable is efficiently implementable.
Furthermore, we remark that Manny Knill~\cite{Knill98} 
has observed that this observable suffices to
also solve the code equivalence problem.  Since linear codes have
canonical forms we may consider the code equivalence problem to be a
hidden stabilizer problem over the same group $S_n \wr S_2$.
See~\cite{PR97} for a discussion of
the relationship of the classical complexities of graph isomorphism and
code equivalence.  

Finally we remark on the group $S_n \wr S_2$
with which we have been working. 
We could equally well work over the subgroup $G^\prime$ 
which is generated by the involutive swaps.  It is
not difficult to show that $G^\prime$ consists of all elements of $G$
of the form $(\sigma, \tau, b)$ where {\em both\/} $\sigma$ and $\tau$
are even or {\em both\/} are odd.  Thus $G^\prime$ has index~2 in $G$ 
and this allows us to work in a smaller Hilbert space.

\section{Acknowledgements}
We would like to thank 
Manny Knill and Richard Hughes, Gian-Carlo Rota and Alain Tapp
for helpful discussions on this problem.

\end{document}